\documentclass[%
,preprint%
,secnumarabic%
,amssymb, amsmath%
, pre%
]{revtex4}
\usepackage{times}
\usepackage{bm}%
\usepackage[%
dvips]{graphicx}

\expandafter\ifx\csname package@font\endcsname\relax\else
 \expandafter\expandafter
 \expandafter\usepackage
 \expandafter\expandafter
 \expandafter{\csname package@font\endcsname}%
\fi

\newcommand{\DegC}{\mbox{\rm\char'27\kern-.3em C}}
\newcommand{\dx}{\partial_x}
\newcommand{\dy}{\partial_y}
\newcommand{\dz}{\partial_z}
\newcommand{\dt}{\partial_t}
\newcommand{\sech}{\mathop{\mathrm{sech}}}
\newcommand{\cm}[1]{({\small \sf #1})} 
\renewcommand\cm[1]{\ignorespaces} 

\begin{document}

\title{Observation of stable phase jump lines in
  convection of a twisted nematic}%
\author{Soichi Tatsumi}
\email{tatsumi@daisy.phys.s.u-tokyo.ac.jp}
\affiliation{Department of Physics, The University of Tokyo, 7-3-1
  Hongo, Bunkyo-ku Tokyo,113-0033}
\author{A. G. Rossberg}
\affiliation{Graduate School of Environment and Information Sciences, 
Yokohama National University, 79-7 Tokiwadai, Hodogaya-Ku
Yokohama 240-8501 }
\author{Masaki Sano}\affiliation{Department of Physics, The University
  of Tokyo, 7-3-1 Hongo, Bunkyo-ku Tokyo,113-0033}

\begin{abstract}
  We report observations of stable, localized, line-like structures
  in the spatially periodic pattern formed by nematic
  electroconvection, along which the phase of the pattern jumps by
  $\pi$.  With increasing electric voltage, these lines form a grid-like 
  structure that goes over into a structure indistinguishable from the
  well known grid pattern.  We present theoretical arguments that
  suggest that the twisted cell geometry we are using is indirectly
  stabilizing the phase jump lines, and that the PJL lattice is caused
  by an interaction of phase jump lines and a zig-zag instability of
  the surrounding pattern.
\end{abstract}
\maketitle

\section{Introduction}
Usually, topological defects in the phase of periodic spatial
modulations have co-dimension two.  They are point-like in two
dimensions, line-like in three dimensions, and in one dimension they
reduce, for dynamical systems, to short events, localized in space and
time\cite{Bray994}.  Here we report the observation of stable, self-organized phase
defects that extend to line-like objects in 2d: \textit{Phase jump
  lines} (PJL) stretched parallel to the wavevector of the modulation,
characterized by a change of the phase of the spatial modulation by
$\pi$ on a short distance (Fig.\ref{SetUp}).

The phenomenon was observed in the periodic pattern of convection
rolls forming in thin layers of the nematic liquid crystal MBBA
(4-methoxybenzylidene-4'-butylaniline) under the influence of an
electric ac field ($\parallel \!\!\hat z$) that is oriented normal to
the ($x$-$y$) layer and the alignment of the nematic director.
Electroconvection (EC) in nematic liquid crystals has been studied
extensively.  After early studies by E. Dubois-Violette and others
\cite{dubdusol} aimed at understanding the basic mechanism of
convection, EC later became popular as a model for pattern formation
in anisotropic systems.  Depending on the experimental conditions, the
convection rolls are aligned either preferentially normal to the
symmetry axis ($\hat x$) of the system---defined by the direction at
which the nematic director is anchored at the boundaries---or oblique
at some preferred angel.  In the oblique-roll case, two convective
modes related to each other by a reflection at the $x$ axis coexist
\cite{zk,bzk}.  Near the onset of convection, descriptions in terms of
one (normal rolls) or two (oblique rolls) time-dependent
Ginzburg-Landau equations could be established \cite{annrev}.  These
equations are of potential type and predict a relaxation to a simple
equilibrium state.

The more puzzling was the rich dynamics of the system observed for
stronger electric fields, with scenarios such as those named
``fluctuating Williams domains'', ``chevron patterns'', ``abnormal
rolls'', ``grid patterns'', or ``defect lattices'' being reported
\cite{kck,asr,huhiroka,RokrChev,Nasuno989-a,Sano992,Nasuno995,Dennin998,Ribotta986,Scherer002,Rudroff999,Sasa990,Oikawa004}.
Only in recent years it became clear that most of these observations can
be understood by the non-relaxational interaction of the convective
mode(s) with only one additional, weakly damped mode that describes an
in-plane rotation of the nematic director. A Ginzburg-Landau-type
model for the non-relaxational coupling of these two (or three) modes,
called \textit{standard model} below, explains most of the experimentaly
observed scenarios at least qualitatively, and predicts a few more
\cite{roth,Komineas003}.

There are several indications in the literature that hint at the
existence of stable PJL in nematic electroconvection.  For example,
for the defect-chaotic regime of normal rolls called ``fluctuating
Williams domains'' it is well known that the ``shape'' of topological
defects, particularly near the events of defect creations and
annihilations, can be highly anisotropic.  Their extension in $x$
direction is considerably larger than in $y$ direction.  As such this
deformation could be explained by the anisotropy of the linear
dispersion relation of the convective mode, which can be removed by a
simple rescaling of the $x$ axis.  But in some cases the defects
stretch considerably and are then more naturally characterized as
``phase jump lines'' \cite{Sasa992,Buka001,Cladis_book001}.  These
structures seem to be related to weakly unstable saddles corresponding
to the stable PJL reported here.  
Stable PJL have also been seen in numerical simulations of the
standard model \cite{zhao00:_secon_instab_compl_patter_anisot_convec}.
Yet, a general theory is missing, and it remains unclear if or under
which conditions stable PJL would be observable in electroconvection.

In our experiments, the conventional EC setup was modified by twisting
the nematic within the layer: Instead of anchoring the nematic
director at the surfaces of the layer parallel to the $x$ axis, the
anchoring direction is rotated by a fixed angle in opposite directions
at the two surfaces \cite{Hertrich994,bosta,Rossberg000,Delev000}.
This geometry seems to be favorable for PJL.

In the following we first describe the details of the twisted EC
setup, and describe the observed PJL and structures they form.  We
then discuss some theoretical ideas that understanding these
observations.

\section{Experimental Setup}
\begin{figure}[bth]
 \includegraphics[width=.9\linewidth]{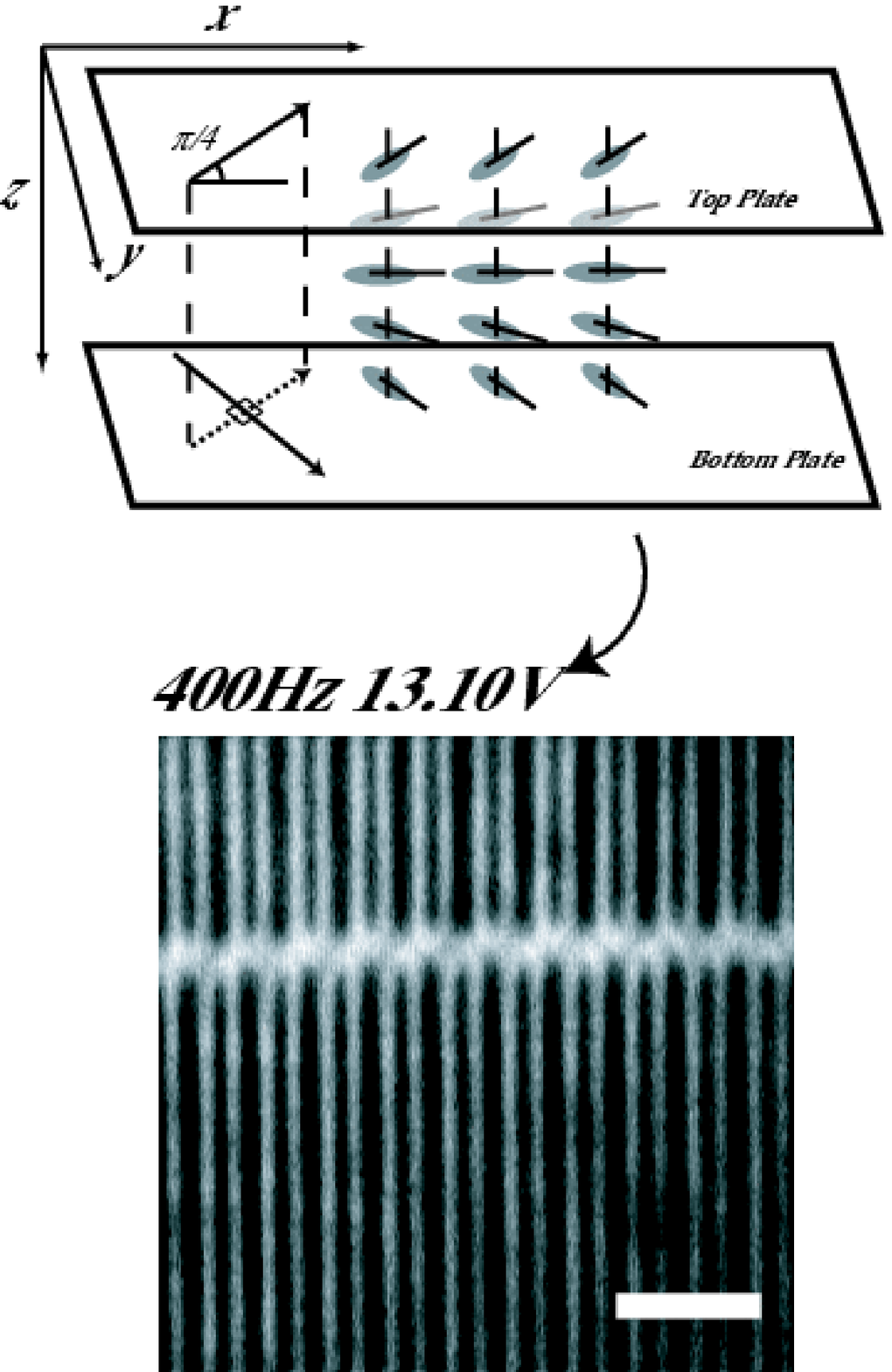}
 \caption{Top: rubbing direction and the anchoring direction of
   nematic. Bottom: a typical phase jump line at 400Hz, 13.10 Volts.
   The white bar corresponds to 100$\mu m$.}
 \label{SetUp}
 \vspace{-.5cm}
\end{figure}
The nematic liquid crystal MBBA doped with 0.1wt\%
tetra-n-butylammonium bromide to stabilize the conductivity was
sandwiched between two glass plates covered with transparent indium
tin oxide thin-film electrodes. The lateral extensions of the liquid
crystal layer were $2\mathrm{cm} \times 2\mathrm{cm}$ and its
thickness $d=50 \mu \mathrm{m}$.  The surfaces of the plates were
coated with polyvinyl alcohol and rubbed in order to attain a planar
anchoring of the nematic molecules.  Temperature of the cell was
controlled at $25\pm0.01 \DegC $. A standard wave-generator was
used to apply ac voltages $V=\mathcal{O}(100\mathrm{V})$ of
frequencies $f$ up to several $\mathrm{kHz}$ to the electrodes.  The
structure of the cell is shown schematically in Fig.~\ref{SetUp}.  The
anchoring direction for each plate is represented by an arrow.  By
rotating the anchoring direction by an angle $\alpha=\pi/4$ out of the
$x$ axis at the lower plate, and in opposite direction by the same
amount at the upper plate, the nematic director is twisted along the
$z$-axis in the ground state \footnote{This geometry allows two
  equivalent ground states: left twist and right twist.  For the
  experiments a region of the sample without domain walls (uniform
  twist) was selected.}  (Fig.~\ref{SetUp}).

\section{Experimental Results}
\begin{figure}[bth]
 \includegraphics[width=\linewidth]{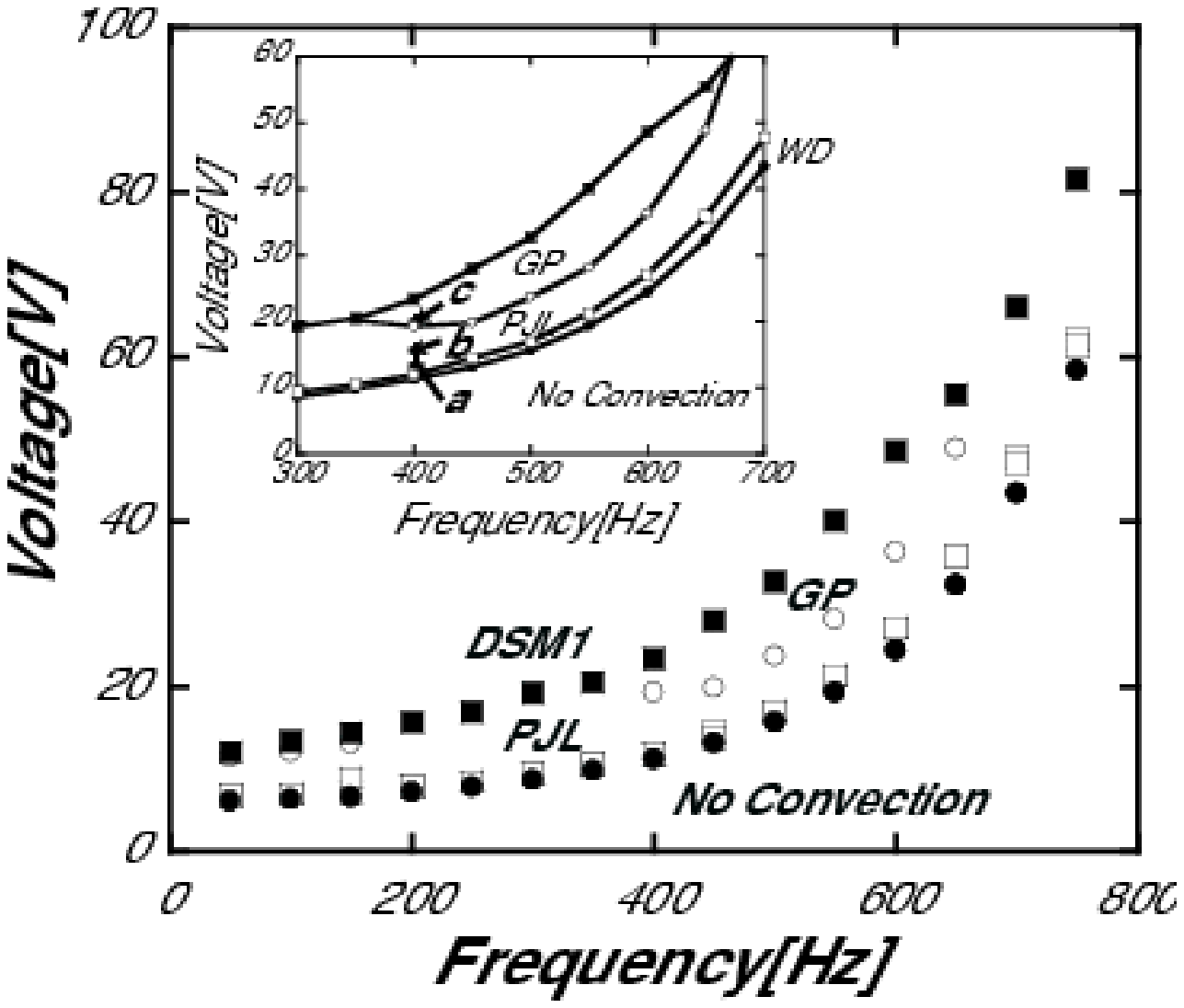}
 \caption{Phase Diagram of our experiments. Solid circles represent the
   onset voltage of convection rolls (Williams Domain); open square
   the onset voltage of PJL; open circle represents the onset of the
   PJL grid pattern (GP); solid square the onset voltage of the
   dynamic scattering mode (DSM1).  The inset shows the part of the
   phase diagram that we discuss in this paper.  The points a, b and c
   in the inset correspond to Figs.~\ref{Main}a-c.
   \label{phase}}
\end{figure}
The control parameters of the system are the externally applied
voltage and its frequency.  A phase diagram of the observed patterns
is shown in the Fig.~\ref{phase}.  The cutoff frequency $f_c$, where
the conductive mode of electroconvection is superseded by the
dielectric mode, was above $1\,\mathrm{kHz}$.  Measurements were
performed for $f<f_c$, {\it i.e.}, in the conductive regime.  As the
voltage was increased, normal rolls with a wavevector parallel to the
$x$-axis (at an angle of $\alpha$ to each anchoring direction) were
observed.  Besides the well-known states including normal rolls
(parallel rolls), grid pattern (rectangular cells), and dynamic
scattering modes (3D turbulent states), we found stably aligned phase
jump lines (PJL) in a wide frequency range of the conductive regime.

\begin{figure*}[bth]
 \includegraphics[width=\linewidth]{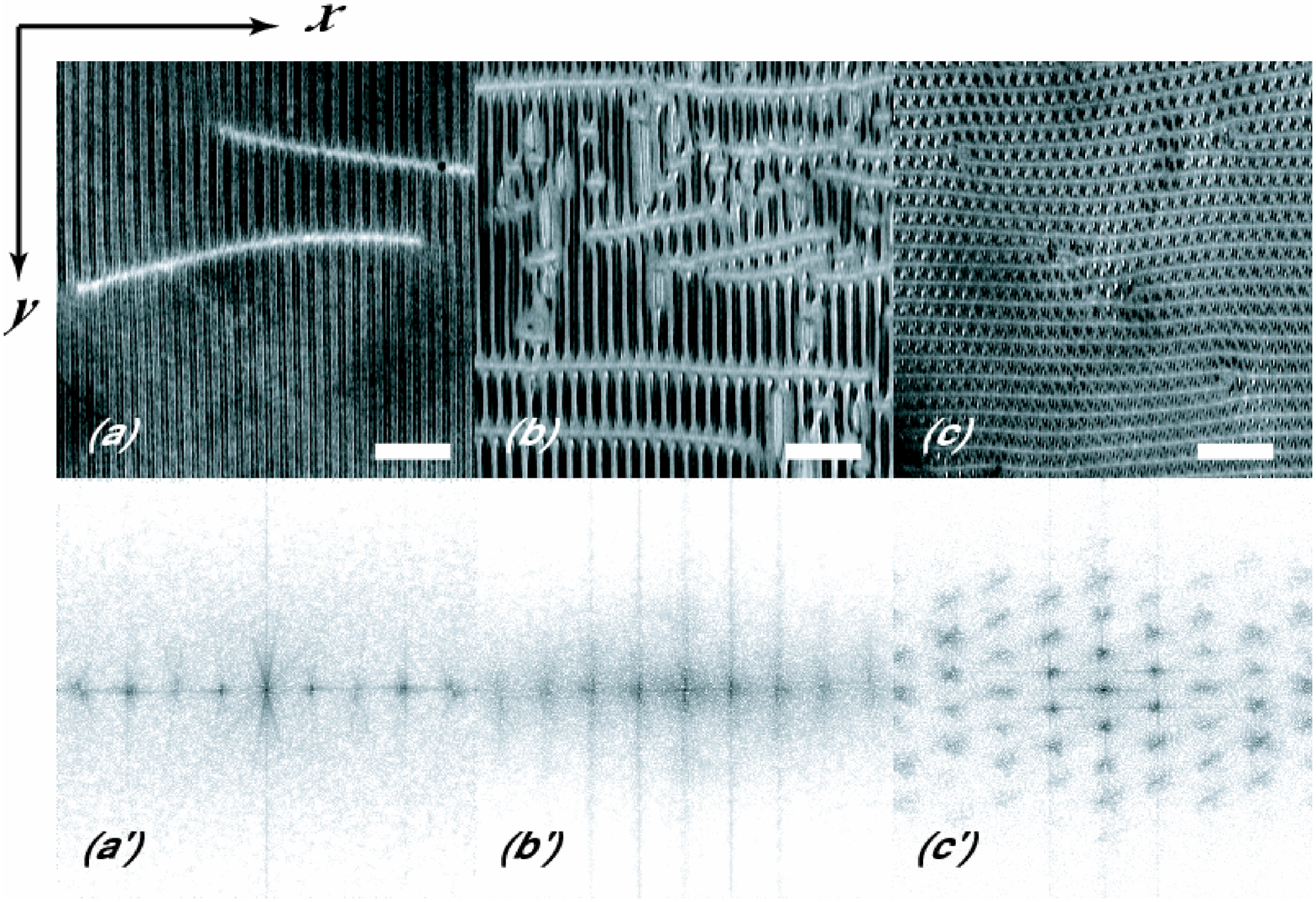}
 \caption{\label{Main} The upper row shows images of convection
   patterns with PJL, the lower row the modulus of the corresponding
   2D Fourier transforms.  These three pair of figures represent the
   transition from PJL to GP.  Fig.~a corresponds to 13.93 V
   ($\epsilon=0.47$), Fig.~b to 15.91 V ($\epsilon=0.91$), and Fig.~c to
   19.23 V ($\epsilon=1.79$), all at 400 Hz. The length of the scale
   bars is 200$\mu m$. }
\end{figure*}
Figures~\ref{Main}a-c show the shadow-graph image of PJL observed
under a polarization microscope (polars $\parallel x$).  PJL can have
finite length.  At the endpoints of PJL the phase difference between
the adjacent domains gradually decreases to zero.  As the voltage is
increased, the typical distance $L$ between PJL in $y$ direction
decreases.  

In order to determine $L$, which is below expressed in terms of the
wave number $K:=2\pi/L$, quantitatively, two methods were employed.
The first makes direct use of the Fourier transforms of recorded
images (Figs.~\ref{Main}b',c'). The other method is to count the
number $n$ \footnote{The number $n$ averaged over a full image was
  determined by first summing the $x$-extension of all PJL in the
  image and dividing this number by the $x$-width of the image.}
of PJL along a line of length $E$ parallel to $\hat y$, which gives
$K=2\pi n/E$.  For high voltage, the Fourier analysis was preferred.
At lower voltages we used the counting method, since the small number
of PJL would not lead to sharp peaks in the Fourier transform.  We
verified that in the intermediate range the results of the two methods
coincide.

\begin{figure}[bth]
 \includegraphics[width=\linewidth]{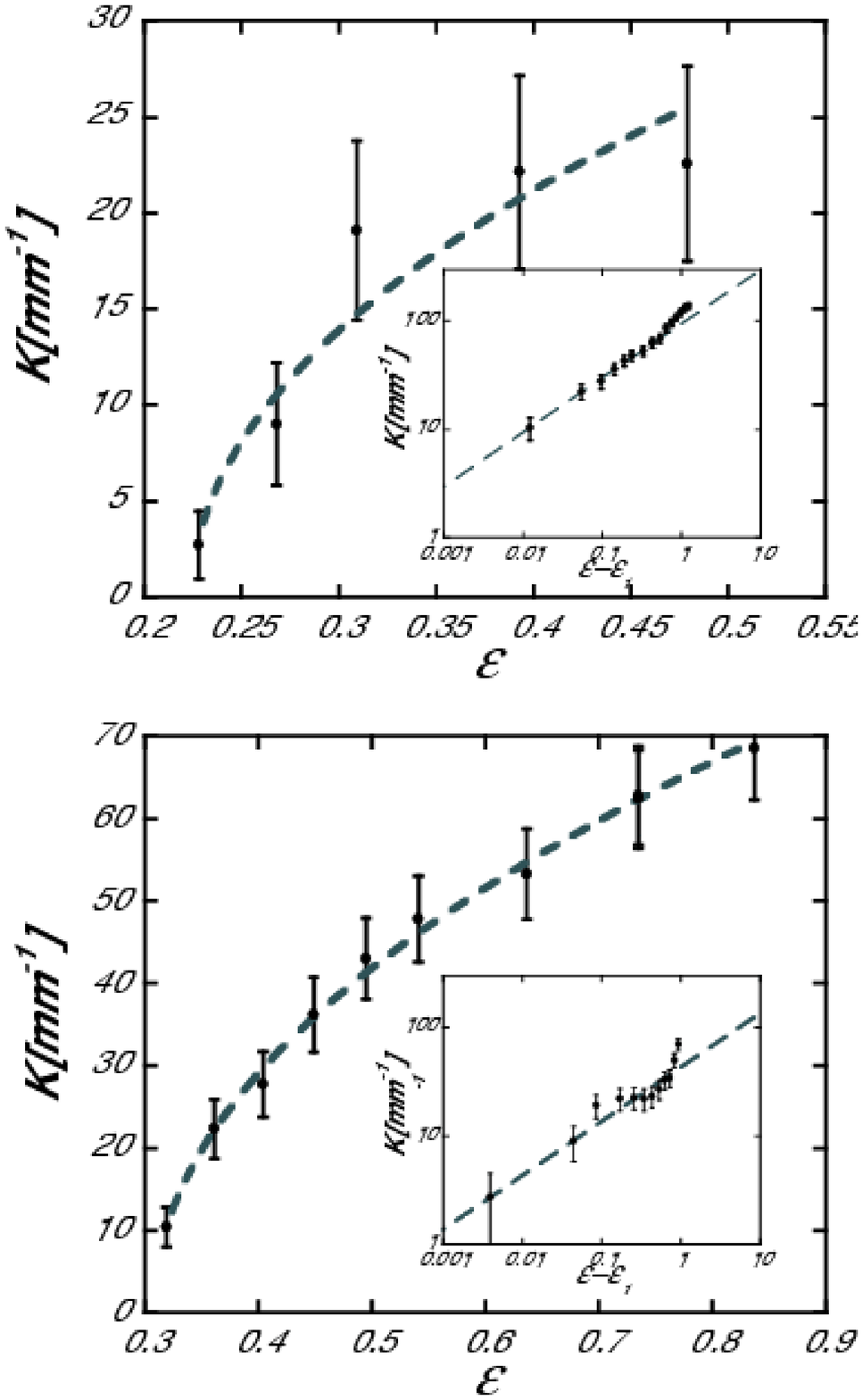}
 \caption{The relation between $\epsilon$ and the inverse distance $K$
   between PJL.  The upper panel corresponds to $f=400$~Hz, the lower
   one to $f=600$~Hz.  The dashed lines are fits to the square-route
   law~(\ref{Klaw}).  The insets show $\epsilon $ - $\epsilon _1$
   \textit{vs} $K$ on a log-log scale.  The onsets of convection is at
   11.50 Volts for 400Hz and at 21.41 Volts for 600Hz.\label{length}
   }
\end{figure}

In Fig.~\ref{length} the measured wave number $K$ is plotted against
the normalized distance $\epsilon = (V^2 -V_c^2)/V_c^2$ from the
threshold of convection $V_c$.  The data indicate that $K$ goes to
zero following a square-root law 
\begin{align}
  \label{Klaw}
 K= c\, (\epsilon - \epsilon_1)^{1/2} 
\end{align}
with $\epsilon_1>0$.  For $f=400\mathrm{Hz}$ we obtain
$c=50.5\mathrm{mm^{-1}}$ and $\epsilon_1=0.22$ and for $f=600Hz$
data is well described by $c=95.3\mathrm{mm^{-1}}$ and
$\epsilon_1=0.31$.

Some observations indicate that there are at least two different
regimes of PJL.  At low $\epsilon$, the spacing between PJL is very
irregular (Fig.~\ref{Main}a,b) and the PJL continue to move, although
very slowly.  At high $\epsilon$, where the spacing between PJL is
smaller, it becomes also more regular (as can be seen from the sharp
peaks in the Fourier transform of Fig.~\ref{Main}c shown in
Fig.~\ref{Main}c'), and a steady state is reached.  We refer to the
resulting structure as a \textit{PJL grid}. The onset voltage of PJL
grids is well above the onset of PJL (Fig.~\ref{phase}).  At this
point the $K$ \textit{vs} $\epsilon$ relation also starts to deviate
significantly from the square-root law valid for smaller $\epsilon$,
as shown in Fig.~\ref{length}.  The deviation is towards larger $K$,
i.e.\ towards denser PJL, contrary to what one would expect from a
repulsive interaction between PJL. The PJL grids are visually
indistinguishable from the well-known grid pattern\cite{Kai976}
consisting of two overlapping sets of oblique
rolls.  If the two structures are indeed identical this means that we
found a new route from normal rolls to oblique rolls, different from
the conventional transition \textit{via} a zig-zag instability.  But
further investigations are required to confirm this hypothesis.  For
the low-$\epsilon$ range, on the other hand, some first steps towards
a better understanding have been made.

\section{Theory}
A clear theoretical picture of PJL, including the conditions for their
existence as well a description of their interaction with the
surrounding roll pattern and with each other, has not yet emerged.
Even though other mechanisms leading to stable PJL appear to exist
\cite{softmodesX}, the simulations of Zhao
\cite{zhao00:_secon_instab_compl_patter_anisot_convec} strongly
suggest that the observed PJL can be described by the standard model
for normal-roll electroconvection.  We shall here investigate three
aspects related to this problem that help  understanding why PJL
stabilize in twisted cells and how they are forming PJL grids.

The standard model \cite{rohekrpe,roth,rzkr,zth,Komineas003} is given by
\begin{subequations}
  \label{Aphi-dim}
  \begin{align}
    \begin{split}
    \label{Aphi-A-dim}
    \tau \partial_t A =& i \beta A \dy \phi + \Big(\epsilon +
    \xi_x^2 \partial_x^2 + \xi_y^2 \partial_y^2 \\ & + 2iq_c\,\xi_y^2\,C_1
    \phi\dy -C_2q_c^2\xi_y^2\phi^2 -g |A|^2\Big) A,
  \end{split}
\\
    \label{Aphi-phi-dim}
    \begin{split}
      \gamma_1 \partial_t \phi =& -h \phi + K_3 \partial_x^2 \phi +
      K_1 \partial_y^2 \phi \\&+2 \Gamma \mathop{\mathrm{Im}}\{A^*
      (\partial_y - i q_c \phi)A\}.
  \end{split}
\end{align}
\end{subequations}
It describes the interaction of two spatio-temporally slow variables:
The complex amplitude $A$ of the convection pattern $\sim
(Ae^{iq_cx}+\text{c.c.})$, and the amount $\phi$ of in-plane rotation
of the nematic director.  By the requirements of structural stability
and according to theoretical and experimental results for the
untwisted geometry, the time-constants $\tau$ and $\gamma_1$, the
coherence lengths $\xi_x$, $\xi_y$, the critical wave number $q_c$,
the geometry factors $C_1$ and $C_2$, the Landau coefficient $g$, the
damping constant $h$, and the two elastic constants $K_1$ and $K_3$
are all positive.  The sign of $\beta$ depends on the applied
frequency and the experimental geometry and is difficult to determine
\textit{a priori}.  The value of $\Gamma$ is always found to be
negative and we shall assume this hereafter.  It is easily seen that
with negative $\Gamma$ convection rolls ($A \ne 0$) have a
destabilizing effect on $\phi$.  This effect is crucial for most of
the complex patterns observed in electroconvection, and PJL are no
exception.

\subsection{Stabilization of PJL}
\label{sec:stability}

After scaling out the natural $\epsilon$ dependence of $A$, $\phi$ and
length and time scales ($\dx\sim\dy\sim A\sim\phi\sim\epsilon^{1/2}$,
$\dt\sim\epsilon$), the standard model~(\ref{Aphi-dim}) can be
reformulated by the following, equivalent system of PDE:
\begin{subequations}
  \label{Aphi}
  \begin{align}
    \label{Aphi-A}
    \tau \partial_t A =& (1 + \partial_x^2 + \partial_y^2 + 2i\,C_1
    \phi\dy-C_2\phi^2 -|A|^2) A + i \beta A \dy \phi, \\
    \label{Aphi-phi}
    \partial_t \phi =& -h\, \phi + \partial_y^2 \phi +
    K_3 \partial_x^2 \phi -2 |\Gamma| \mathop{\mathrm{Im}}\{A^* (\partial_y - i
    \phi)A\}.
\end{align}
\end{subequations}
The coefficients and variables in system~(\ref{Aphi}) are not
identical to those denoted with the same symbols in
system~(\ref{Aphi-dim}), but are related to these by simple scaling
transformations.  For all coefficients but $h$ this relation is
independent of $\epsilon$.  The coefficient $h$ in
Eq.~(\ref{Aphi-phi}) scales as $\epsilon^{-1}$, i.e., $h$ plays the
role of the main control parameter in the rescaled
equations~(\ref{Aphi}).  As $h$ decreases, the first instability of
the homogeneous solution $A=1$, $\phi=0$ is to abnormal rolls at
$h=2|\Gamma|$ if $\beta<0$ and to zig-zag modulations at
$h=2|\Gamma|(1+\beta)$ if $\beta>0$.

The system also has a steady-state solution
\begin{align}
  \label{PJL}
  \phi=0, \quad A=\tanh
  \left(
     y/\sqrt{2}
  \right)
\end{align}
that describes PJL.  For large $h$, i.e.\ close to the threshold of
convection, PJL are unstable.  This can be seen by considering the
limit $h\to\infty$ where $\phi=0$ is fixed.  Then solution~(\ref{PJL})
is unstable to the linear mode $\delta A=i\,\sech( y/\sqrt{2})$ with
growth rate $\tau^{-1}$ \cite{langer967}.  However, by lowering $h$
the additional degree of freedom $\phi$ can stabilize the PJL by an
interaction through the last term in Eq.~(\ref{Aphi-phi}).  For a
discussion of this process, we linearize the system~(\ref{Aphi}) with
respect to perturbations of solution~(\ref{PJL}) of the form $\delta
A= [a_r(y)+i a_i(y)]\exp(\sigma t)$, $\delta \phi=f(y) \exp(\sigma
t)$, which leads to
\begin{align}
  \label{ar}
  0 =& \left( -\sigma\,\tau +1
    -3\,{{\tanh}^2}{\frac{y}{{\sqrt{2}}}}+\partial_y^2 \right) a_r
  ,
\end{align}
and
\begin{subequations}
\label{linearized}
\begin{align}
  \label{ai}
  \begin{split}
    0 =& \left(- {\sqrt{2}}\,{ C_1}\,
      {{\sech}^2}{\frac{y}{{\sqrt{2}}}} +
      \beta\,\tanh{\frac{y}{{\sqrt{2}}}}\,\partial_y \right) f
    \\
    & + \left( -\sigma\,\tau + {{\sech}^2}{\frac{y}{{\sqrt{2}}}}
      +\partial_y^2 \right) a_i,
\end{split}
  \\
  \label{delta}
  \begin{split}
    0 =& \left(-\sigma-h -
      2|\Gamma|{{\tanh}^2}{\frac{y}{{\sqrt{2}}}}+\partial_y^2\right) 
    f \\
    & - \left({\sqrt{2}}\,|\Gamma|
      {{\sech}^2}{\frac{y}{{\sqrt{2}}}}-
      2\,|\Gamma|\,\tanh{\frac{y}{{\sqrt{2}}}}\partial_y \right)
    a_i .
\end{split}
\end{align}
\end{subequations}
We are looking for eigenfunctions of these equations which are bounded
for $y\to\pm\infty$.  Equation~(\ref{ar}) is decoupled from
Eqs.~(\ref{ai},\ref{delta}).  Apart from the translational mode
$a_r=\dy \tanh(y/\sqrt{2})$, Eq.~(\ref{ar}) has bounded solutions only
with $\tau \sigma\le-3/2$ and does not contribute to the linear
stability problem.  The $y$ axis can be divided into an inner region
$|y|\lesssim 5$ near the PJL, and two outer regions where the
system~(\ref{linearized}) simplifies.  For $y>0$, for example, the
substitutions $\sech\to 0$, $\tanh\to 1$ lead to
\begin{subequations}
  \label{outer}
  \begin{align}
    \label{ai-out}
    0=&(-\sigma\tau+\dy^2) a_i+\beta \dy f,\\
    \label{delta-out}
    0=&(-\sigma+2|\Gamma|\mu+\dy^2)f-2 |\Gamma| \dy a_i,
  \end{align}
\end{subequations}
where we introduced the abbreviation $\mu=1-h/2|\Gamma|$.
In general, the eigenmodes of Eq.~(\ref{linearized}) have to be
calculated numerically.  For example this can be done by ``shooting''
solutions starting at $y=0$ with $a_i=1$, $f=f_0$, $a_i'=f'=0$ into
the $y>0$ outer region such as to match
$a_i(y_1),a_i'(y_1),f(y_1),f'(y_1)$  for
sufficiently large $y_1$ with a bounded outer solution.  
We shall here only be interested in the
cases where the PJL stabilize prior to any destabilization of the
homogeneous state, i.e., for the cases that $\mu<-\beta$ and $\mu<0$.  The
bounded outer solutions are then generally given by linear
combinations of the two solutions of~(\ref{outer}) of the form
$a_i,f\sim\exp(\lambda y)$ with $\lambda\le 0$ and
\begin{align}
  \label{flat}
  \lambda^2=&\frac{\mu\sigma\tau}{\beta+\mu}+\mathcal{O}(\sigma^2)
  \intertext{or}
  \label{steep}
  \lambda^2=&-2|\Gamma|\,(\beta+\mu)+\mathcal{O}(\sigma),
\end{align}
respectively.  The value of $\mu$ where the PJL stabilize can be found
by considering the limit $\sigma\to 0^+$ of the
systems~(\ref{linearized}) and~(\ref{outer}).  Two particularities of
this limit will be used here: (i)~The outer solution corresponding
to~(\ref{flat}) goes over into a simple phase shift ($f=0$,
$a_i=\text{const}.$).  Thus, $f$ contains only contributions from the
component corresponding to~(\ref{steep}) and must satisfy
\begin{align}
  \label{phi-steep}
  \dy f=-\sqrt{-2|\Gamma|(\beta+\mu)}\,f.
\end{align}
(ii)~By Eq.~(\ref{ai-out}) the quantity
\begin{align}
  \label{conserved}
  u=\dy a_i+\beta f\quad(=0)
\end{align}
is conserved along $y$ in the outer region
\cite{zhao00:_secon_instab_compl_patter_anisot_convec}.  When
$\mu<-\beta$ and~(\ref{phi-steep}) holds, this implies $u=0$ for
bounded solutions.  Equations~(\ref{phi-steep},\ref{conserved}) are
necessary and sufficient conditions for the inner solutions to be
bounded when continued to $y\to\infty$.  When the shooting method is
used, the two conditions can be satisfied by adjusting the initial
value $f_0$ of $f$ at $y=0$ and the value of the control parameter
$\mu$.

\begin{figure}[tbp]
  \centering
  \includegraphics[width=\linewidth,keepaspectratio,clip]{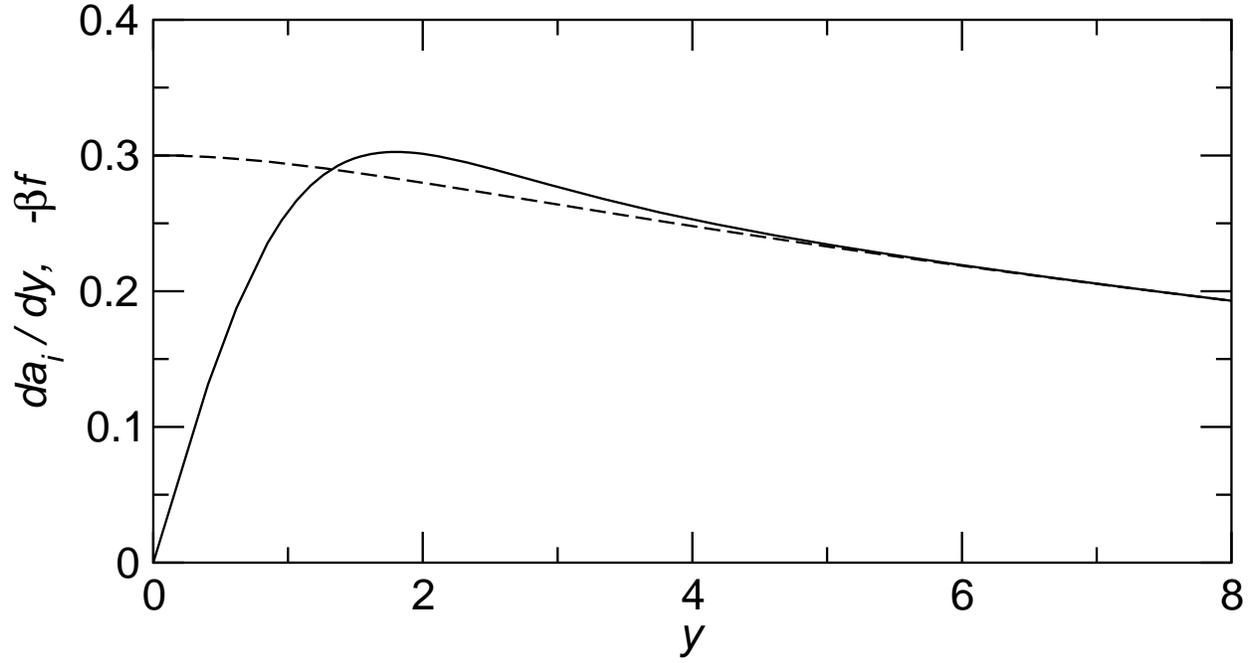}
  \caption{Typical neutral mode of a phase jump line at the critical
    control parameter. Solid: $\dy a_i$, dashed: $-\beta f$.}
  \label{fig:stable-mode}
\end{figure}

With $C_1>1$ we could find neutrally stable modes of PJL with
$\mu<-\beta$ for all parameter sets tested, in particular for positive
as well as negative $\beta$.  For example,
Fig.~(\ref{fig:stable-mode}) shows the neutrally stable mode of the
PJL for parameters $\Gamma=-0.6$, $\beta=0.3$, $C_1=1.1$, and
$\tau=0.5$.  These parameters are suggested by numerical calculations
for similar systems~\cite{rohekrpe}.  For this case we verified that
there is no unstable mode, i.e., Fig.~(\ref{fig:stable-mode}) shows
the critical mode of a PJL.  (For smaller $|\Gamma|$, e.g. with
$\Gamma=-0.06$, there is another mode with positive growth rate.  Then
the feed-back loop through $\phi$ is too weak to stabilized the PJL.)

As $C_1$ is lowered, the critical $\mu$ approaches $-\beta$ and the
decay of the $f$ component in the far field becomes flatter, as
described by Eq.~(\ref{phi-steep}). Thus $a_i$ increases/decreases
approximately linearly over a wide range according
to~(\ref{conserved}).  For $\mu\to-\beta$, one expects the critical
mode to converge to an unbounded secular solution with constant $\dy
a_i$ and $f$ in the outer range.

It turns out that for $C_1=1$ the linearized
equations~(\ref{linearized}) have the analytic solution
\begin{align}
  \label{analytic-mode}
  f=1,\quad a_i=\sqrt{2}\,(1-\mu)+\mu\, y\, \tanh\frac{y}{\sqrt{2}}.
\end{align}
For $\mu=-\beta$ this solution satisfies conditions~(\ref{phi-steep})
and (\ref{conserved}) and is of the expected form.  Combined with our
numerical calculations, we conclude that the critical $\mu$ reaches
$-\beta$ exactly when $C_1=1$.  Thus, for $\beta_y>0$, the PJL can
stabilize prior to the zig-zag instability at $\mu=-\beta$ only if
$C_1>1$.  
%

For $\beta<0$ the first instability of the homogeneous state and the
far field in system~(\ref{linearized}) is towards abnormal rolls at $\mu=0$.  
In contrast to the zig-zag instability at $\mu=-\beta$, the outer
equations~(\ref{outer}) have for $\mu=0$ no secular solutions with
$\sigma=0$.  Instead, the solution space corresponding to $\lambda=0$
[Eq.~(\ref{flat})] is spanned by two linearly independent constant
solutions, the pure phase shift $(a_i,f)=(1,0)$ and the pure director
rotation $(a_i,f)=(0,1)$.  Since both are bounded, bounded solutions
of~(\ref{linearized}) for $\mu=0$ are obtained whenever the
contribution to the outer solutions corresponding to Eq.~(\ref{steep})
with $\lambda>0$ vanishes.  When the shooting method is used, this can
be achieved by adjusting a single parameter, e.g. $f_0$.  Thus, at the
abnormal-roll instability $\mu=0$, a neutrally stable linear mode of
the PJL can always be found.

By the interaction of $f$ and $a_i$ in the core region, a certain
linear combination of the constant outer modes is singled out,
characterized by a specific ratio $r=a_i/f$ (for $y\to+\infty$).  When
$C_1=1$, for example, the analytic solution~(\ref{analytic-mode})
gives $r=\sqrt{2}$.  A perturbative analysis of the outer
equations~(\ref{linearized}) for small $\sigma\sim\dy\sim\mu$ under
the constraint $a_i/f=r+\mathcal{O}(\mu)$ yields a solution with
$a_i,f\sim\exp(\lambda_s y)$ (as $y\to+\infty$) where
\begin{subequations}
\label{muexpand}
\begin{align}
  \label{lambdas}
  \lambda_s=&\frac{2 r \tau |\Gamma| \mu}{2 r^2 |\Gamma|
    \tau-\beta}+O(\mu^2) 
  \intertext{with the growth rate}
  \label{sigmas}
  \sigma=&\frac{2 \beta |\Gamma| \mu}{2 r^2 |\Gamma|
    \tau-\beta}+O(\mu^2).
\end{align}
\end{subequations}
When $\lambda_s<0$ this solution is bounded and an eigenmode of the
PJL.  Since $\sigma/\lambda_s=\beta/r\tau+\mathcal{O}(\mu^2)$, this
mode can be stable only for $r<0$, which is the case only for values
of $C_1$ much smaller or much larger than one.  Usually one would
assume $r>0$.  With $-2|\Gamma|\, r^2\,\tau<\beta<0$, for example,
(\ref{muexpand}) then describe an unstable linear mode that becomes
neutrally stable as $\mu$ approaches zero from below.

In conclusion, these considerations show that within the framework of
the model~(\ref{Aphi}) a stabilization of PJL of the form~(\ref{PJL})
prior to any destabilization of the homogeneous state---as we seem to
observe it in the experiments---is possible, but likely only for
$C_1>1$ and $\beta\gtrsim 0$. 

\subsection{The effect of the twisted anchoring on $C_1$}
\label{sec:c1}
In order to test if the twisted anchoring used in our experiment can
cause a higher value of $C_1$, we analyzed a model for the $z$
dependence of director twist $\hat \phi(z)$ and convection mode $\hat
A(z)$ and their interaction \cite{softmodesX,twisted}.  Keeping only
terms that are relevant for determining $C_1$, the model reads in
appropriate dimensionless units
\begin{subequations}
  \label{zequations}
  \begin{align}
    \label{zA}
    \tau \dt \hat A=&\left[u-(P-\hat \phi)^2+\dz^2 \right] \hat A\\
    \label{zphi}
    \dt \hat \phi=&\dz^2 \hat \phi + 2\Gamma |\hat A|^2 (P-\hat \phi).
  \end{align}
\end{subequations}
The coefficient $u$ in Eq.~(\ref{zA}) is a control parameter that
depends on the applied voltage. $P$ corresponds to the angle of the
wavevector of a plane-wave convection mode to the $x$ axis and the
term $-(P-\hat \phi)^2$ in Eq.~(\ref{zA}) describes the preference of
the convective modes to align normal to the director.  The last term
($\dz^2$) describes the diffusion of hydrodynamic variables along $z$.
In Eq.~(\ref{zphi}) the first term describes the stiffness of the
in-plane director to twist and the second term the torque generated by
misaligned rolls on the director.  We assume layers of thickness $2$
without loss of generality and impose the boundary conditions $\hat
A(\pm 1)=0$ and, corresponding the twisted anchoring, $\hat
\phi(+1)=b$, $\hat \phi(-1)=-b$.  The isotropy of the system is broken
only by the boundary condition.  In the bulk, isotropy is preserved,
which is reflected by the equivariance under $P\to P+\delta$, $\hat
\phi\to\hat \phi+\delta$.  The model can be derived analytically in
the limit that the wavelength of the roll pattern is much smaller than
the layer thickness \cite{softmodesX}.  However, since this condition
is not satisfied for our experimental system, we expect the model to
describe the experiment only qualitatively.

The coefficient $C_1$ entering the 2d description~(\ref{Aphi}) can be
obtained by the following method.  Noting that in the basic state
($\hat A=0$) the director relaxes to $\hat \phi=b z$, and that the
fundamental mode of linear excitations of $\hat \phi$ is $\hat
\phi_1=\cos(\pi z/2)$, we assume a weakly excited $\hat \phi=b z+K
\hat \phi_1$ and search numerically for the threshold value
$u_\text{crit}$ of the control parameter $u$ at which the lowest
eigenmode $\hat A(z)=\hat A_\text{crit}(z)$ of Eq.~(\ref{zA}) becomes
critical ($\dt=0$).  The orientation $P$ of the roll wavevector is
chosen such as to minimize $u_\text{crit}$.  This corresponds to
letting $A=\exp(iPy)$ in Eq.~(\ref{Aphi-A}) with $P=C_1 \phi$.
Following general prescriptions (e.g.\ \cite{manbook,bkbook}), the
value of $C_1$ is then given by the ratio of projections of two
components of the r.h.s.\ of Eq.~(\ref{zphi}) onto its adjoint
eigenvector ($\equiv \hat \phi_1$)
\begin{align}
  \label{c1}
  C_1=\frac{\displaystyle \int_{-1}^1 \hat \phi_1 \, 2\Gamma |\hat A|^2 P\, dz}
  {\displaystyle \int_{-1}^1 \hat \phi_1 \, 2\Gamma |\hat A|^2 \hat \phi\, dz}
\end{align}
in the limit $K \to 0$.  

\begin{figure}[bth]
  \centering
  \includegraphics[width=\linewidth,keepaspectratio,clip]{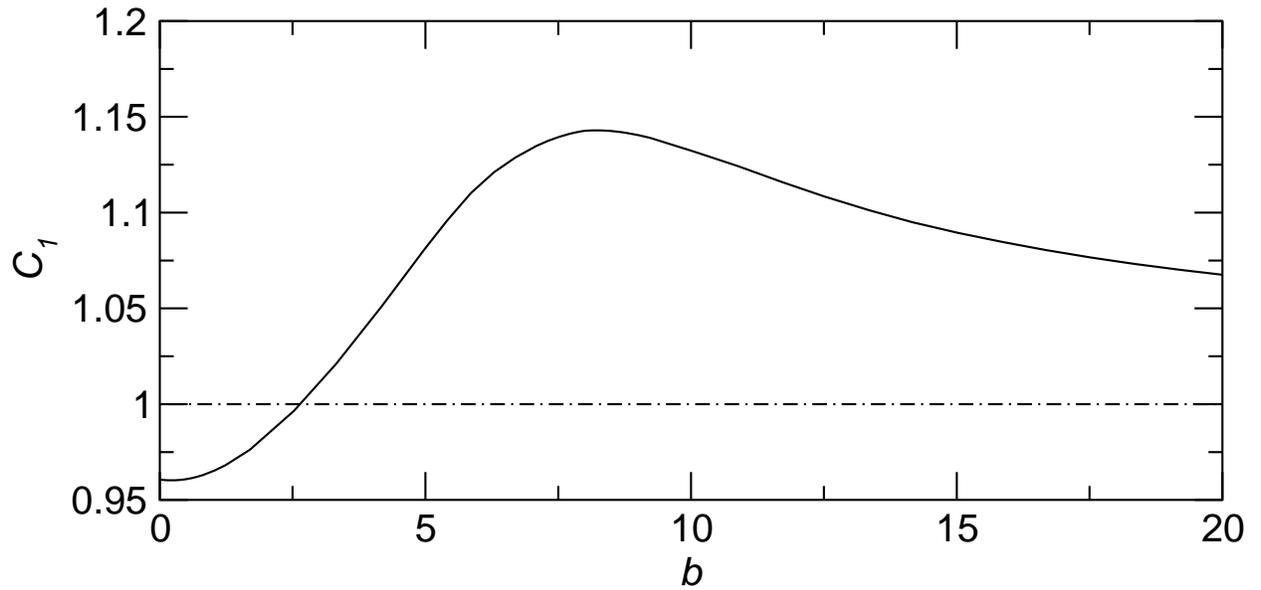}
  \caption{The geometry factor $C_1$ in dependence of the twist $b$
  for the simple model~(\ref{zequations}).  The dash-dotted line at
  $C_1= 1$ is a guide to the eye.}
  \label{fig:c1}
\end{figure}

Without twist ($b=0$) we reproduce the analytic result \footnote{This
  result does not depend on the absence or presence of
  flexoelectric effects, as a note in Ref.~\cite{Komineas003} might
  suggest.}  $C_1=256/(27\pi^2)=0.96$ \cite{linddpl,softmodesX}.  With
increasing twist $b$, the value of $C_1$ increases, until it reaches a
maximum $C_1=1.14$ at $b=8.1$ (Fig.~\ref{fig:c1}).  The result is
independent of all other system parameters.  Due to the purely
geometric nature of this effect, it is plausible to assume that the
twist leads to an increased value of $C_1$ also in our experimental
system.

\subsection{Coupling to zig-zag modulations}
\label{sec:zig-zag}
Simulations~\cite{zhao00:_secon_instab_compl_patter_anisot_convec}
show that PJL can also coexist with zig-zag modulations of the roll
pattern.  The PJL are then located at preferred phase angles of the
zig-zag modulation.  Thus, the distance $L$ between neighbouring PJL
will be controlled by the wavelength $\Lambda$ of the zig-zag
modulation.  Even though they are not clearly visible in the
experiments, small zig-zag modulations are probably also controlling
the distance between PJL in PJL grids.

To test this hypothesis, the wavelength of zig-zags is estimated.  For
an easy comparison with experiments, dimensional units are used in
this calculation.  In the absences of strong variations of the pattern
amplitude $|A|$, the dynamics of the roll pattern can be described in
terms of the phase $\theta=\arg A\,(\mathop\mathrm{mod}2\pi)$ of the
pattern alone.  Together with the coupling to $\phi$, this is
description becomes \cite{rzkr,Zhao000}
\begin{subequations}
  \label{thetaphi}
  \begin{align}
    \label{thetaphi-theta}
    \tau\dt \theta=&
    \xi_x^2\dx^2\theta+\xi_y^2\dy^2\theta+\beta \dy \phi+r\phi^2\dy\phi, \\
    \label{thetaphi-phi}
    \gamma_1\dt\phi=&
    2|\Gamma|(\epsilon-\epsilon_\text{AR})\phi-g\phi^3+K_3\dx^2\phi+
    K_1\dy^2\phi-2|\Gamma|\epsilon\dy\theta
  \end{align}
\end{subequations}
with positive $r,g$, the threshold of abnormal rolls
$\epsilon_\text{AR}=h/(2|\Gamma|)$ and all other parameters identical
to those entering Eq.~(\ref{Aphi-dim}).  While the derivation of the
nonlinear terms in this system hinges on the assumption of that
spatial and temporal variations of $\theta$ and $\phi$ are slow on the
scales $\xi_{x,y}\epsilon^{-1}$ and $\tau\epsilon^{-1}$, which is
strictly satisfied only for small $\epsilon-\epsilon_\text{AR}$ and
$\beta$, the linear part is valid uniformly in $\epsilon\le0$ and
$\beta$, as long as the Ginzburg-Landau description (\ref{Aphi-dim})
is valid.

For stationary solutions that depend only on $y$ ($\dt,\dx=0$)
Eq.~(\ref{thetaphi-theta}) can be integrated to
\begin{align}
  \label{theta-integrated}
  \xi_y^2\dy \theta+\beta \phi+r\phi^3/3=J=\text{const.}
\end{align}
In the simple case that rolls are not tilted on the average ($\left<
  \dy\theta \right>_y=0$) the integration constant $J$ vanishes.
Using Eq.~(\ref{theta-integrated}) to eliminate $\theta$ from
Eq.~(\ref{thetaphi-phi}) then yields an equation for $\phi$ of the
form~\cite{Zhao000}
\begin{align}
  \label{phi-alone}
  K_1\dy^2\phi=-2 |\Gamma|((1+\beta/\xi_y^2)
  \epsilon-\epsilon_\text{AR})\phi+g'(\epsilon)\phi^3.
\end{align}
While the wavelength of zig-zag modulations can become arbitrary large
when the nonlinearity in Eq.~(\ref{phi-alone}) becomes effective, the
wavelength of small-amplitude modulations is sharply determined as
\begin{align}
  \label{wavelength1}
  \begin{split}
    \Lambda=&2\pi \sqrt{\frac{K_1}{2|\Gamma|
        ((1+\beta/\xi_y^2)\epsilon-\epsilon_\text{AR})}}\\
=&2\pi \sqrt{\frac{K_1
        \epsilon_\text{ZZ}}{2|\Gamma|\epsilon_\text{AR}
        (\epsilon-\epsilon_\text{ZZ})}}.
\end{split}
\end{align}
where $\epsilon_\text{ZZ}=\epsilon_\text{AR}/(1+\beta/\xi_y^2)$
denotes the threshold of the zig-zag instability.  In order to
estimate the value of $2|\Gamma|\epsilon_\text{AR}/K_1$, recall that
the linear part of Eq.~(\ref{thetaphi-phi}) holds also for
$\epsilon=0$ (no convection).  In this case the relaxation of small
in-plane rotations of the director is easily described by the
linearized Leslie-Erickson \cm{spell?} equations.  For the untwisted
case, for example, they reduce to
\begin{align}
  \label{LE}
  \gamma_1\dt n_y=(k_{33}\dx^2+k_{11}\dy^2+k_{22}\dz^2) n_y.
\end{align}
Considering only the slowest mode of the system for fixed boundary
conditions at $z=\pm d/2$, $n_y=\phi \cos(\pi z/d)$, and projecting
onto the corresponding adjoint $\cos(\pi z/d)$, yields
\begin{align}
  \label{rederived}
  \gamma_1\dt \phi=(k_{33}\dx^2+k_{11}\dy^2-k_{22}\frac{\pi^2}{d^2}) \phi.
\end{align}
Comparing of Eqs.~(\ref{thetaphi-phi}) and (\ref{rederived}) shows
that $2|\Gamma|\epsilon_\text{AR}/K_1=\pi^2 k_{22}/d^2 k_{11}$.  For
MBBA at 25$^\circ$C, for example, the ratio $k_{11}/k_{22}$ is
$\approx 1.6$.  For the twisted geometry one would expect some mixture
of $k_{11}$, $k_{22}$, and $k_{33}$ to enter $K_1$ instead of just
$k_{11}$, but since in the projection onto the adjoined mode the
central part $z\approx 0$ dominates, and there the director is still
aligned parallel to $\hat x$, these corrections are presumably small.
Assuming a relation $L=m\,\Lambda$ between zig-zag wavelength
$\Lambda$ and PJL distance $L$, one obtains the result
\begin{align}
  \label{L}
 \begin{split}
  L=&2 d m\sqrt{\frac{k_{11}
  \epsilon_{ZZ}}{k_{22}(\epsilon-\epsilon_{ZZ})}}. \\
  \Leftrightarrow
  K=&\frac{\pi}{dm}\sqrt{\frac{k_{22}}{k_{11}}\frac{\epsilon-\epsilon_{ZZ}}{\epsilon_{ZZ}}}. 
 \end{split}
\end{align}
Even when $\epsilon_\text{ZZ}$ is fitted from the measured data for
$K$, Eq.~(\ref{L}) makes a nontrivial prediction because
$\epsilon_\text{ZZ}$ appears at two different positions in the
formula.

Using the fitted experimental curves, we find that Eq.~(\ref{L}) holds
with $m=2.1$ for the $400\,\mathrm{Hz}$ data and with $m=0.94$ for the
$600\,\mathrm{Hz}$ data.  That is, in both cases we find a value of
$m$ of order one.  In view of the coarseness of the theoretical
estimates, and allowing for some experimental error, we conclude that
the observations fully supports the hypothesis that the lattice
constant of the PJL-grid is controlled by small zig-zag modulations.
The mechanism by which the amplitude of zig-zag modulations or,
equivalently, the zig-zag wavelength is kept small remains unclear,
however.

\section{Conclusion}
In this study, we reported the observation of stable PJL and described
a process by which these PJL from a lattice that finally evolves to a
structure indistinguishable from the well know grid pattern as the
applied voltage increases.  We argued that the stabilization of PJL is
presumably an indirect effect of the twisted geometry of the nematic
in the basic state: The twist leads to values of the coefficient
$C_1>1$, which in turn enables the stabilization of PJL prior to a
destabilization of the roll pattern.

\bibliographystyle{apsrev}
\bibliography{bibview1_3,bibtatsumi} 

\end{document}